\documentclass[a4paper,11pt]{article}
\usepackage{jcappub} 

\usepackage{changes}

\usepackage{amsmath} 
\usepackage{xcolor}    
\usepackage{cancel}
\setdeletedmarkup{\ifmmode\cancel{#1}\else\sout{#1}\fi}

\arxivnumber{2505.19055} 
\title{\boldmath  Testing the Cosmological Principle on Gigaparsec Scales}

\author[a]{Xin Wang}
\author[a,b,1]{and Zhiqi Huang\note{The corresponding author.}}
\affiliation[a]{School of Physics and Astronomy, Sun Yat-sen University,\\
Zhuhai, 519082, P.R.China}
\affiliation[b]{CSST Science Center for the Guangdong-Hongkong-Macau Greater Bay Area,Sun Yat-sen University,\\
Zhuhai, 519082, P.R.China}

\emailAdd{huangzhq25@mail.sysu.edu.cn}
\emailAdd{wangx898@mail2.sysu.edu.cn}

\abstract{Recent observational analyses have suggested possible evidence of hemisphere asymmetry in cosmological datasets. Parameterizations of this kind place observers in a privileged position—specifically on the plane that divides the two hemispheres. To quantify potential deviations from the cosmological principle without presuming a special location, we develop a stochastic framework that parametrizes departures from statistical homogeneity and isotropy. The near-uniform temperature of the cosmic microwave background indicates that anisotropy is negligible (at the $\lesssim 10^{-5}$ level) on the last scattering surface. This serves as a zero boundary condition, enabling the construction of an orthogonal basis of functions below the recombination redshift. Within this basis, we expand the relative deviation from the Hubble diagram of isotropic models (such as $\Lambda$CDM or $w_0w_a$CDM) in a hierarchy of increasing resolution. Applying this approach, we test the cosmological principle using Type Ia supernovae, strong lensing time delays, and gravitational-wave standard sirens. For the class of large-scale anisotropies and low-order radial variations described by this framework, the current datasets are found to be consistent with statistical homogeneity and isotropy on gigaparsec scales.}

\def\aap{Astronomy \& Astrophysics}

\def\apjl{Astrophysical Journal Letters}
\def\mnras{MNRAS}
\def\jcap{JCAP}
\def\prd{Physical Review D}
\def\prl{Physical Review Letters}
\def\apj{Astrophysical Journal}

\def\nat{Nature}
\def\araa{Annual Review of Astronomy and Astrophysics}

\begin{document}

\maketitle
\flushbottom

\section{Introduction}

The development of modern cosmology is largely based on the cosmological principle, which states that the universe is homogeneous and isotropic on large scales~\cite{Milne_1932, Aluri:2022hzs}. As an extension of the Copernican principle on cosmological scales, the cosmological principle was initially based more on philosophical considerations than rigorous observational evidence. The evolution of modern cosmology has led to a quantitative refinement of the cosmological principle. Within the standard $\Lambda$CDM model—whose parameters have been precisely constrained by cosmic microwave background (CMB) observations—the low-redshift universe is expected to be approximately homogeneous and isotropic on scales larger than $100\,\mathrm{Mpc}$.  Observational verification of the cosmological principle on scales $\gtrsim 100\,\mathrm{Mpc}$ constitutes a fundamental validation of the cornerstone of the modern standard cosmological model, carrying profound implications for our understanding of cosmic structure formation.

The empirical support of the cosmological principle dates back to the 1960s, when astronomers began mapping the sky in microwave and radio bands \cite{Penzias_1965, Wilson_1967, Partridge_1967, Ryle_1968}. While early evidence remained largely qualitative, advances in recent decades have enabled more rigorous statistical tests, primarily through studies of extragalactic objects in the CMB rest frame \cite{Ellis_1984}. Multi-probe observations have since revealed several anomalies that may indicate a violation of the cosmological principle. These include a dipole excess in radio sources \cite{Singal_2011, Rubart_2013, Tiwari_2016, Bengaly_2018, Singal_2019, Siewert_2021, Singal_2023} and quasars \cite{Secrest_2021, Secrest_2022, Dam_2023, Wagenveld_2024}, as well as unexpectedly large bulk flows inferred from low-redshift galaxies \cite{Watkins:2008hf, Feldman:2009es, Watkins:2023rll} and supernovae \cite{Sorrenti_2023, Tang_2023, Sah_2024}. More recently, signatures of a quadrupole component in supernovae~\cite{Cowell_2023, Dhawan_2023, Sorrenti_2025} as well as anomalous signatures in other various forms~\cite{Shurtleff:2013sua,Pelgrims_2015,Shamir_2025, Huang_2022b} have been reported.

When investigating high-redshift ($z\sim 1$) regimes of the cosmos, where bulk flow measurements become increasingly observationally prohibitive, SNe serve as critical diagnostic tools for probing the cosmological principle on larger scales. Nevertheless, whether high-redshift SNe exhibit statistical isotropy remains contentious, with conflicting conclusions across studies and no definitive consensus to date~\cite{Deng:2018jrp, Zhao:2019azy, Hu_2024, Hu:2024qnx, Yang_2025}. The most pronounced anisotropy on large scales has been found with a hemisphere comparison method, where the discrepancy between cosmological parameters fitted in two hemispheres is maximized by varying the direction of hemisphere splitting~\cite{Schwarz_2007, Antoniou_2010, Cai:2011xs, Yang:2013gea, Zhao:2019azy, Hu:2020mzd, Conville_2023, Hu:2024qnx}. However, even within the hemisphere comparison framework, the statistical significance of cosmological principle violation remains critically sensitive to the choice of parameters and no consensus has been reached. For example, by fitting the Hubble constant $H_0$ in hemispheres, Ref.~\cite{Hu:2024qnx} finds a preferred direction at $4.39\sigma$ significance level, while hemisphere comparison of the deceleration parameter $q_0$ only yields an $1.78\sigma$ significance which can be well interpreted as a statistical fluctuation. 

Despite its widespread application, the hemisphere comparison method may introduce three potential problems. First, it places the observer in a privileged position - specifically on the plane that divides the two hemispheres, thereby violating the Copernican principle a priori. Second, the hemisphere splitting of cosmological parameters may contradict the observed almost uniform temperature of the CMB. Third, parameters chosen for hemisphere comparison, such as the Hubble constant used in Ref.~\cite{Hu:2024qnx}, may be only sensitive to the local data, while their interpretation risks being erroneously generalized to imply global hemispheric asymmetry. 

To address these limitations, we propose a new approach that incorporates a full three-dimensional expansion of the Hubble residuals. The CMB last scattering surface serves as a natural boundary where Hubble residuals are known to be negligible ($\lesssim 10^{-5}$). We decompose the Hubble residuals into orthonormal basis functions constructed with zero boundary condition on the last scattering surface, thereby ensuring consistency with CMB isotropy constraints. Finally, the radial dependence of the basis functions separates the contributions from data at different redshifts, thereby avoiding misinterpreting a local anisotropy as a global one. 

This paper is organized as follows. Section~\ref{sec:method} gives the detailed description of our method and the data sets that we use. Results are given in Section~\ref{sec:res}. Section~\ref{sec:con} discusses and concludes. The matter abundance parameter $\Omega_m$ is defined as the ratio between the current background matter density and the critical density $\frac{3H_0^2}{8\pi G}$, where $G$ is the Newton's gravitational constant. The right ascension (RA) and declination (DEC) are defined in the International Celestial Reference System (ICRS), while spherical harmonics are evaluated in the Galactic coordinate frame. 

\section{Method \label{sec:method}}

A point-like object at redshift $z$ and in direction $\mathbf{n}$ is labeled with a reference coordinate $\left(\tilde{r}(z), \mathbf{n}\right)$, where $\tilde{r}(z)$ is the comoving distance in a reference isotropic model, such as $\Lambda$CDM. The true comoving distance of the object, which we allow to be anisotropic, is decomposed into orthonormal functions in the reference coordinate space,
\begin{equation}
  r(z, \mathbf{n}) = \tilde{r}(z) \left[1+\sum_{i=1}^{i_{\max}}\sum_{\ell=1}^{\ell_{\max}}\sum_{m=-\ell}^\ell C_{i\ell m}\,\mathrm{j}_{\ell}\left(\frac{\mu_{\ell i}\tilde{r}(z)}{\tilde{r}(z_{\rm rec })}\right)\mathcal{Y}_{\ell m}\left(\mathbf{n}\right)\right],     \label{eq:model}
\end{equation}
where $\mu_{\ell i}$ is the $i$-th positive root of the spherical Bessel function $\mathrm{j}_\ell$. The real spherical harmonic functions  $\mathcal{Y}_{\ell m}$  is defined as
\begin{equation}
\mathcal{Y}_{\ell m} \equiv
\begin{cases} 
\frac{\mathrm{Y}_{\ell m} + (-1)^m \mathrm{Y}_{\ell,-m}}{\sqrt{2}}, & m > 0; \\
\mathrm{Y}_{\ell m}, & m = 0; \\
\frac{\mathrm{Y}_{\ell,-m} - (-1)^m \mathrm{Y}_{\ell m}}{\sqrt{2}i}, & m < 0,
\end{cases}
\end{equation}
where $\mathrm{Y}_{\ell m}$ is the spherical harmonic function. Given the accurately measured CMB temperature $T_{\rm CMB}\approx 2.726\,\mathrm{K}$~\cite{Fixsen_2009}, the recombination redshift $z_{\rm rec}\approx 1089$ is well determined in the standard picture of recombination~\cite{Seager_1999}, and is therefore fixed in our calculation. By construction, $r(z_{\rm rec}, \mathbf{n})$ is identical to $ \tilde{r}(z_{\rm rec})$ in all directions, thereby satisfying the CMB isotropy constraint.

\begin{figure}
  \centering
  \includegraphics[width=\linewidth]{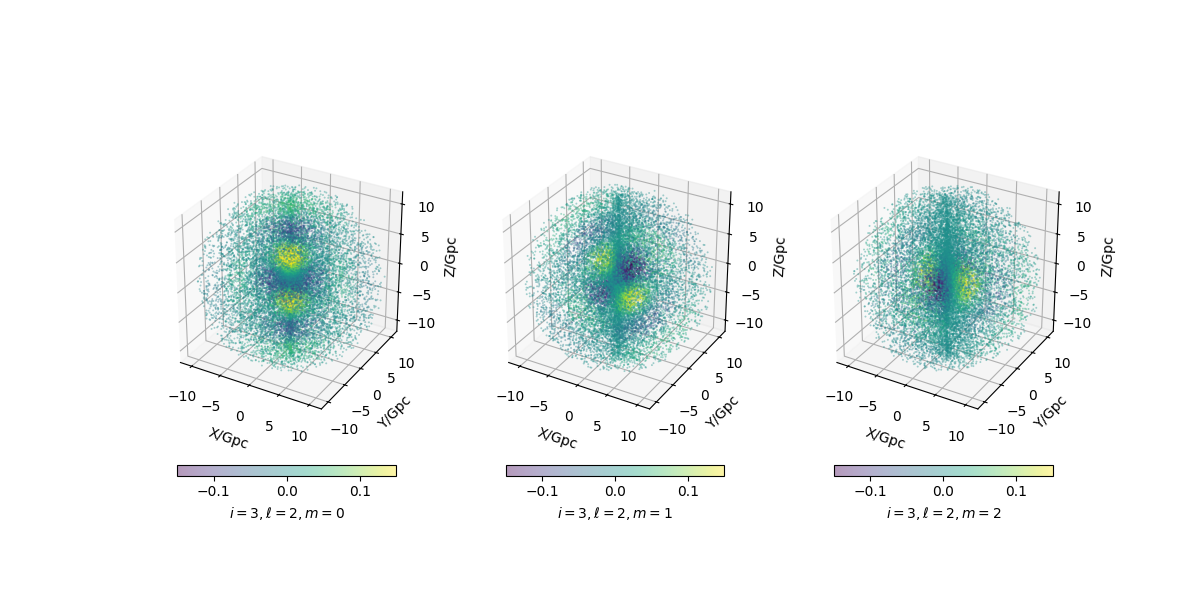}
  \caption{Expansion basis $\mathrm{j}_{\ell}\left(\frac{\mu_{\ell i}\tilde{r}(z)}{\tilde{r}(z_{\rm rec })}\right)\mathcal{Y}_{\ell m}\left(\mathbf{n}\right)$  in comoving coordinate, with $\ell=2, i=3$ and from left to right, $m=0, 1, 2$.\label{fig:basis}}
\end{figure}

Although an arbitrary function can, in theory, be expanded at infinite spatial resolution ($\ell_{\max} = i_{\max} = \infty$), practical data limitations constrain us to a finite set of a few thousand points. This necessitates truncating the expansion at finite values of the angular and radial indices, $\ell_{\max}$ and $i_{\max}$. Figure~\ref{fig:basis} illustrates several typical basis functions in comoving coordinates. A truncation with $\ell_{\max}$ and $i_{\max}$ on the order of a few corresponds to a spatial resolution on gigaparsec scales. For reference, most of the data lie within the range of $z \in [0, 1.5]$, corresponding to a comoving distance of $r \in [0, 4.5,\mathrm{Gpc}]$.

We consider three isotropic reference models: $\Lambda$CDM, $w_0w_a$CDM, and the PAge approximation. The $w_0w_a$CDM model extends $\Lambda$CDM by allowing a dynamical dark energy equation of state, parameterized as $w(a) = w_0 + w_a(1-a)$~\cite{Chevallier:2000qy,Linder:2002et}. In contrast, the PAge parameterization~\cite{Huang_2020, Huang_2021, Huang_2022a, Wang_2024, Su_2025} offers a more model-independent approach by not presupposing a specific two-fluid (dark matter plus dark energy) cosmology. It introduces two parameters: the cosmic age $p_{\rm age}$ in units of $H_0^{-1}$, and $\eta$, which quantifies the deviation from Einstein-de Sitter cosmology.

Since our analysis incorporates the SH0ES distance-ladder measurement, the absolute SN magnitude $M$ is included as a free parameter. We adopt uniform priors for all cosmological parameters: $\Omega_m \in [0.2, 0.4]$, $H_0 \in [60, 90] \ \mathrm{km \cdot s^{-1} Mpc^{-1}}$, $w_0 \in [-3, 1]$, $w_a \in [-3, 2]$, $p_{\rm age} \in [0.8, 1.5]$, $\eta \in [-1, 1]$, and $M \in [-25, -15]$. For the anisotropy coefficients, we use $C_{i\ell m} \in [-5, 5]$ for all $i, \ell, m$. Additionally, we impose two physicality priors: $13 \ \mathrm{Gpc} < \tilde{r}(z_{\rm rec}) < 15 \ \mathrm{Gpc}$ to prevent the derived CMB acoustic peak locations from becoming grossly inconsistent with observations, and $w_0 + w_a < 0$ to ensure negative pressure for dark energy.

The primary data sets we use for this study are the Pantheon+ and SH0ES supernovae~\cite{Brout_2022}, publicly available at \url{https://github.com/PantheonPlusSH0ES/DataRelease}. The cosmological redshift (denoted as $z_{\rm HD}$ in the catalogue file) was obtained by removing the estimated peculiar velocity contribution from the CMB-frame redshift $z_{\rm CMB}$. Data points with $z_{\rm HD} < 0.01$ are excluded in our analysis, as in this regime the observed redshift $z_{\rm HEL}$ (and CMB-frame redshift $z_{\rm CMB}$) may be dominated by peculiar motion, making $z_{\rm HD}$ unreliable due to uncertainties in the peculiar velocity estimates. Several studies have shown that bulk flows up to $z \lesssim 0.1$ may affect anisotropies on scales up to $\sim 250\,\mathrm{Mpc}$~\cite{Mazurenko_2024, Koksbang_2024, Tang_2023, Sorrenti_2023, Sorrenti_2025}. However, in the present study, our spatial resolution is limited to $\mathrm{Gpc}$ scales or larger. Consequently, our results are not expected to be sensitive to such local, bulk-flow-induced anisotropies. Instead, our analysis is better suited to probing potential large-scale violations of the cosmological principle arising from nonstandard frameworks, such as modified gravity or nonstandard initial fluctuations~\cite{Tzartinoglou_2024, Li_2020, Li_2021, Cai_2022, Li_2022, Li_2023, Zhu_2023, Rao_2023, Jiang_2024, Bond_2009, Huang_2019, McInnes_2025, Huang_2025}. Nevertheless, we will conduct a more detailed examination of the impact of peculiar velocity corrections at the end of Section~\ref{sec:res}.

We modify the publicly available likelihood to incorporate the revised anisotropic comoving distance, Eq.~\eqref{eq:model}. Hereafter for brevity we will implicitly refer the selected 1657 SNe as ``Pantheon+'', without specifying the inclusion of SH0ES and the exclusion of the $z_{\rm HD}<0.01$ subset.

\begin{figure}
    \centering
    \includegraphics[width=\linewidth]{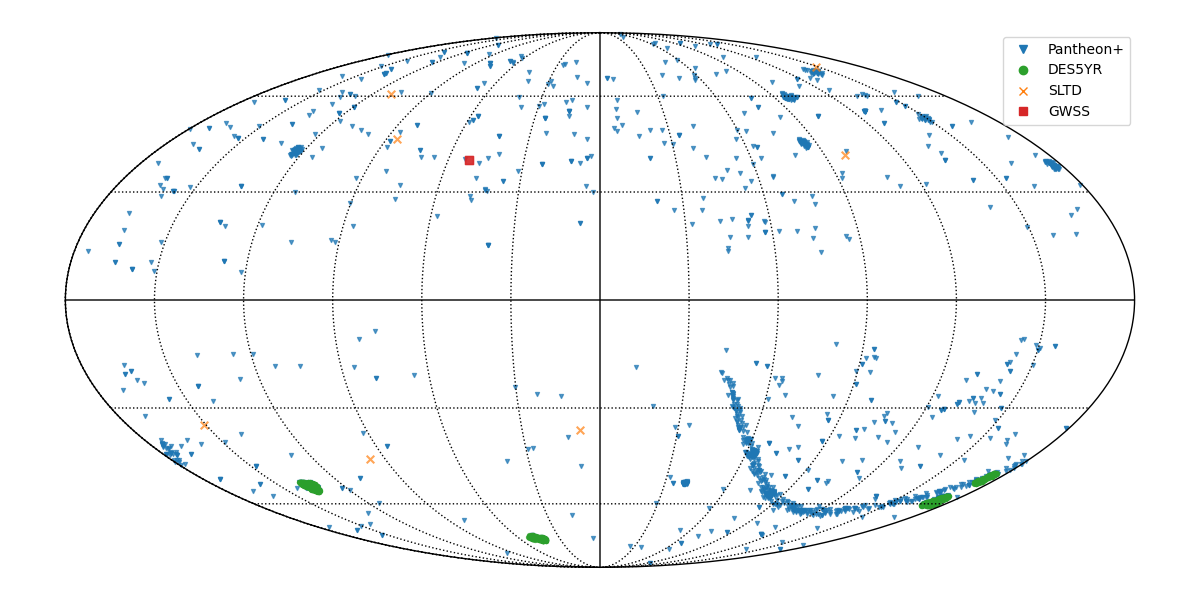}
    \caption{Galactic coordinates of the data.\label{fig:data}}
\end{figure}

To check consistency between Pantheon+ and other cosmological data sets, we optionally include additional multi-probe data that are collectively shown in Figure~\ref{fig:data}. The extra data include seven data points of strong lens time delay (SLTD)~\cite{Wong_2020_H0LiCOW,Shajib_2020_STRIDES}, one data point of gravitational-wave standard siren (GWSS)~\cite{Abbott_2017}, and 1635 SNe from the full five year of the Dark Energy Survey (DES5YR) SN program~\cite{DES_SN_2024}. The summary statistics of SLTD are listed in Table~\ref{tab:SLTD}. The GWSS data is a constraint on the angular diameter distance  $d_A(z=0.01006, \mathrm{RA}=197.4487^\circ, \mathrm{DEC}=-23.3840^\circ) = 43.8^{+2.9}_{-6.9}\,\mathrm{Mpc}$, where $d_A(z,\mathbf{n})=\frac{r(z,\mathbf{n})}{1+z}$. The DES5YR SN data and likelihood are publicly available at \url{https://github.com/des-science/DES-SN5YR}. We also adapted the DES5YR supernova likelihood by removing supernovae that overlap with the Pantheon+ sample and incorporating the anisotropic model given in Eq.~\eqref{eq:model}. The final DES5YR catalogue used in our analysis comprises 1491 supernovae.

\begin{table}
\centering
  \renewcommand{\arraystretch}{1.3}
\caption{Summary statistics of the strong lens time delay data, where $z_l$ and $z_s$ are the redshifts of lenses and sources, respectively. \label{tab:SLTD}}
\begin{tabular}{cccccc}
\hline
\hline
Lens name & RA(deg) & DEC(deg) & $z_l$ & $z_s$ & $\frac{r(z_l)r(z_s)}{r(z_s)-r(z_l)}$(Mpc)\\
\hline
B1608+656\cite{Suyu:2009by,Jee:2019hah} & $242.3081$ & $65.5411$ & $0.6304$ & $1.394$ & $5156^{+296}_{{-236}}$ \\
RXJ1131-1231\cite{Suyu:2013kha,H0LiCOW:2019xdh} & $172.9644$ & $-12.5329$ & $0.295$ & $0.654$ & $2096^{+98}_{-83}$ \\
HE0435-1223\cite{H0LiCOW:2016qrm,H0LiCOW:2019xdh} & $69.5619$ & $-12.28745$ & $0.4546$ & $1.693$ & $2707^{+183}_{-168}$ \\
SDSSJ1206+4332\cite{H0LiCOW:2018tyj} & $181.62361$ & $43.53856$ & $0.745$ & $1.789$ & $5769^{+589}_{-471}$ \\
WFI2033-4723\cite{H0LiCOW:2019mdu} & $308.4257$ & $-47.3956$ & $0.6575$ & $1.662$ & $4784^{+399}_{-248}$ \\
PG1115+080\cite{H0LiCOW:2019xdh} & $169.57035$ & $7.76627$ & $0.311$ & $1.722$ & $1470^{+137}_{-127}$ \\
DESJ0408-5354\cite{Shajib_2020_STRIDES,Agnello:2017mwu} & $62.0905$ & $-53.8999$ & $0.597$ & $2.375$ & $3382^{+146}_{-115}$ \\
\hline
\end{tabular}
\end{table}

We run Markov Chain Monte Carlo (MCMC) simulations to infer cosmological parameters as well as the anisotropy coefficients $C_{i\ell m}$'s. A detection of nonzero $C_{i\ell m}$ typically implies a violation of cosmological principle, though in the high-resolution (large $\ell_{\max}$, $i_{\max}$) case more careful Bayesian model comparison should be done to avoid the over-fitting problem. A more convenient approach is to use the Gaussian approximation to estimate the statistical significance of the preference for anisotropy.
When the posterior of parameters are obtained from Markov Chains, we compute the mean and the covariance matrix of the $C_{i\ell m}$ coefficients, denoted as $v$ and $\mathrm{Cov}$ respectively. The tail probability of the isotropic model (all $C_{i\ell m}=0$) in the Gaussian approximation is given by
\begin{equation}
 p = \frac{2^{1-\frac{N_c}{2}}}{\Gamma\left(\frac{N_c}{2}\right)}\int_{\rho_0}^\infty e^{-\rho^2/2}\rho^{N_c-1}~\mathrm{d}\rho,
\end{equation}
where  $N_c = i_{\max}\sum_{\ell=1}^{\ell_{\max}}(2\ell+1) = i_{\max}\ell_{\max}(\ell_{\max}+2)$ is the total number of the $C_{i\ell m}$ coefficients and $\rho_0\equiv \sqrt{v^{\mathrm{T}}\mathrm{Cov}^{-1}v}$. The preference for anisotropy can also be framed in more intuitive terms by expressing its significance in sigma units. The number of sigmas, $n_\sigma^{\rm aniso}$, is linked to the tail probability via $p = \mathrm{erfc}\,\left(\frac{n_\sigma^{\rm aniso}}{\sqrt{2}}\right)$, where $\mathrm{erfc}$ is the complementary error function.

We also utilize the Akaike Information Criterion (AIC) and the Bayesian Information Criterion (BIC) as complementary measures to the Gaussian approximation-based model comparison. Specifically, AIC is defined as $\mathrm{AIC} = \chi^2_{\min} + 2k$, where $\chi^2_{\min}$ denotes the minimum chi-square statistic and $k$ represents the number of free parameters. Lower AIC values correspond to models that achieve a better trade-off between goodness-of-fit and model complexity, as the penalty term $2k$ discourages overparameterization. Similarly, BIC is defined with a stronger penalty term, $k \ln N_d$, where $N_d$ is the sample size (e.g., $N_d = 1635$ for the Pantheon+ dataset).

\section{Results \label{sec:res}}

Table~\ref{tab:allruns} summarizes the analyses of the Pantheon+ SNe with various reference isotropic models and angular resolutions of anisotropy. At $\lesssim 1.5\sigma$ confidence level, the data allow all the anisotropy coefficients $C_{i\ell m}$'s to be zero simultaneously. Isotropic models with significantly lower AIC and BIC values are favored by the data. Thus, up to the resolution we have tested, Pantheon+ is in excellent agreement with cosmological principle. 

Among the isotropic models, $\Lambda$CDM has the lowest BIC and PAge has the lowest AIC. No strong preference for either of the isotropic models should be claimed, as the differences between the AICs of the isotropic models are small ($\lesssim 2$).

\begin{table}[htpb]
    \centering
        \caption{Statistical significance of anisotropy in Pantheon+ SNe \label{tab:allruns}}
      \renewcommand{\arraystretch}{1.3}
    \begin{tabular}{ccccccc}
    \hline
    \hline
    reference model & $\ell_{\max}$ & $i_{\max}$  & $n^{\rm aniso}_\sigma$ & $\chi^2_{\min}$ & AIC & BIC \\
    \hline
     $\Lambda$CDM  &  $0$  & $0$ & - & $1455.1$ & $1461.1$ & $1477.3$ \\
       &  $1$ & $1$ &  $0.87$ & $1452.6$ & $1464.6$ & $1497.0$ \\
       &  $1$ & $2$ &  $0.70$ & $1450.2$ & $1468.2$ & $1516.8$ \\
       &  $1$ & $3$ &  $0.39$  & $1449.5$ & $1473.5$ & $1538.3$ \\
      &  $1$ & $4$ &  $0.35$  & $1448.9$ & $1478.9$ & $1559.9$ \\
       & $1$ & $5$ &  $0.44$ &  $1445.7$ & $1481.7$ & $1578.9$ \\        
       &  $2$ & $1$ &  $0.62$ & $1449.5$ & $1471.5$  & $1530.9$ \\
       &  $2$ & $2$ &  $1.15$ & $1437.2$ & $1475.2$ & $1577.7$ \\
       &  $2$ & $3$ &  $0.81$ & $1434.8$ & $1488.8$ & $1634.6$ \\    
      & $3$ & $1$ & $1.29$ & $1435.9$ & $1471.9$ & $1569.1$ \\
    &  $3$ & $2$ &  $1.02$ & $1432.3$ & $1498.3$ & $1676.5$ \\        
       & $4$ & $1$ &  $0.90$ & $1432.6$ & $1486.6$ & $1632.4$ \\    
    
     \hline
     $w_0w_a$CDM & $0$ & $0$ & - & $1451.7$ & $1461.7$ & $1488.7$ \\
      & $1$ & $1$ & $0.34$ & $1450.3$ & $1466.3$ & $1509.5$ \\ 
    \hline
     PAge & $0$ & $0$ & - & $1451.7$ & $1459.7$ & $1481.3$ \\
      & $1$ & $1$ & $0.39$ & $1450.4$ & $1464.4$ & $1502.2$ \\
    \hline
    \end{tabular}
\end{table}

Next, we incorporate the DES5YR supernova sample, SLTD, and GWSS into our analysis. Combining supernova datasets with different calibration procedures and light-curve models is not commonly done in the literature. In particular, DES5YR employs the SALT3 light-curve fitting model~\cite{Kenworthy_2021}, while Pantheon+ uses SALT2~\cite{Guy_2007}. A key difference between the two datasets is characterized by $\Delta \mathcal{M} = \mathcal{M}_{\rm DES5YR} - \mathcal{M}_{\rm Pantheon+}$, which quantifies the difference in standardized SN absolute magnitudes. By comparing the overlapping samples in Pantheon+ and DES5YR, Ref.~\cite{Efstathiou_2025} identified a significant offset in $\Delta\mathcal{M}$ between $z<0.1$ and higher redshifts. According to Ref.~\cite{Vincenzi_2025}, this offset primarily stems from the updated intrinsic scatter model in SALT3 and improved host galaxy stellar mass estimates for certain low-redshift samples. However, the $\Delta\mathcal{M}$ offset does not pose an issue in our analysis, as the overlapping samples present in both Pantheon+ and DES5YR have been excluded from the DES5YR dataset.

\begin{table}[htpb]
    \centering
        \caption{Testing anisotropy with SNe (Pantheon+\&DES5YR) + SLTD + GWSS \label{tab:multiprobe}}
      \renewcommand{\arraystretch}{1.3}
    \begin{tabular}{cccccccc}
    \hline
    \hline
   reference model & $\ell_{\max}$ & $i_{\max}$  & $n^{\rm aniso}_\sigma$ & $\chi^2_{\min}$ & AIC & BIC & $\Delta\mathcal{M}$\\
    \hline
    $\Lambda$CDM  &  $0$  & $0$ & - & $2800.6$ & $2808.6$ &  $2832.8$ & $0.0225\pm 0.0068$\\ 
       &  $1$ & $1$ &  $1.05$ & $2797.8$ & $2811.8$ &  $2854.1$  & $0.0323\pm 0.00091$\\
    &  $1$ & $2$ & $0.46$ & $2794.6$ & $2814.6$ &  $2875.1$ & $0.0312 \pm 0.0091$ \\
    &  $1$  &   $3$ &  $0.63$ &  $2793.5$ &  $2819.5$ &  $2898.1$ &  $0.0321\pm 0.0091$ \\
    &  $1$  &   $4$ &  $0.39$ &  $2793.5$ &  $2825.5$ &  $2922.3$ &  $0.0315\pm 0.0091$ \\
    &  $1$  &   $5$ &  $0.75$ &  $2793.3$ &  $2831.3$ &  $2946.2$ &  $0.0370\pm 0.0169$ \\        
    &  $2$ & $1$ & $1.06$ & $2791.6$ & $2815.6$ &  $2888.2$ & $0.0323\pm 0.0094$ \\
    & $2$ & $2$ & $0.70$ & $2784.8$ & $2824.8$ &  $2945.8$ & $0.0346\pm 0.0097$ \\
    &  $2$  &   $3$ &  $0.61$ &  $2781.9$ &  $2837.9$ &  $3007.3$ &  $0.0359\pm 0.0099$ \\
    & $3$ & $1$ & $1.15$ & $2782.3$ & $2820.3$ & $2935.2$ & $0.0337\pm 0.0099$ \\
    & $4$ & $1$ & $0.88$ & $2777.5$ & $2833.5$ & $3002.8$ & $0.0351\pm 0.0100$ \\    
     \hline
    $w_0w_a$CDM & $0$ & $0$ & - & $2797.4$ & $2809.4$ & $2845.7$ & $0.0337 \pm 0.0087$ \\
      & $1$ & $1$ & $0.40$ & $2796.2$ & $2814.2$ & $2868.6$ & $0.0380\pm 0.0097$ \\ 
      & $2$ & $1$ &  $0.79$ & $2788.9$ & $2816.9$ & $2901.6$ & $0.0402\pm 0.0102$ \\  
     \hline
     PAge & $0$ & $0$ & - & $2797.4$ & $2807.4$ & $2837.6$ & $0.0346\pm 0.0088$  \\
          & $1$ & $1$ & $0.38$ & $2796.2$ & $2818.2$ & $2860.6$ & $0.0385\pm 0.0096$  \\
          & $2$ & $1$ & $0.72$  & $2790.0$   & $2816.0$ & $2894.6$ & $0.0396\pm 0.0100$\\
    \hline
    \end{tabular}
\end{table}

\begin{figure}
    \centering
    \includegraphics[width=\linewidth]{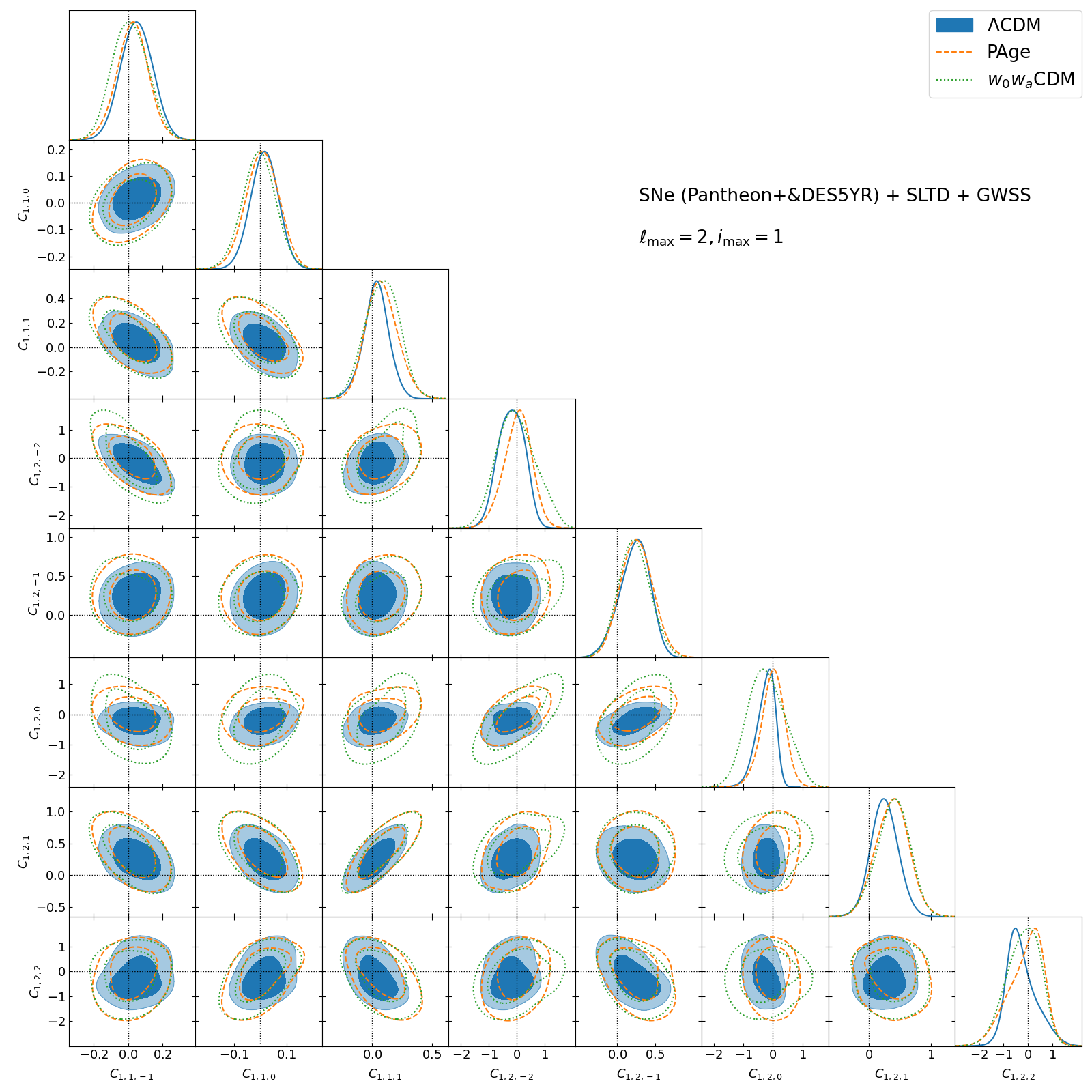}
    \caption{Marginalized 68.3\% and 95.4\% confidence-level constraints on anisotropy coefficients.\label{fig:aniso}}
\end{figure}

Table~\ref{tab:multiprobe} summarizes the results of model comparison, which again favor the isotropic models. Figure~\ref{fig:aniso} shows the marginalized posteriors of the anisotropy coefficients of the $\ell_{\max}=2$, $i_{\max}=1$ case as a typical example. In all cases, the cosmological principle ($C_{i\ell m} = 0, \forall i,\ell, m$) is in excellent agreement with the joint multi-probe data. 

As the last column of Table~\ref{tab:multiprobe} shows, a nonzero $\Delta\mathcal{M}$ is found at $\gtrsim 3\sigma$ confidence level in all cases. The posterior of $\Delta \mathcal{M}$ has a noticeable cosmology dependence. In particular, the mean value of $\Delta\mathcal{M}$ shifts by about $1.5\sigma$ from isotropic $\Lambda$CDM to  models beyond $\Lambda$CDM (anisotropic or $w_0w_a$CDM or PAge). The degeneracy between $\Delta M$ and cosmology indicates that there may still be a trend of redshift evolution of $\Delta\mathcal{M}$ at $z>0.1$. 

\begin{figure}
    \centering
    \includegraphics[width=\linewidth]{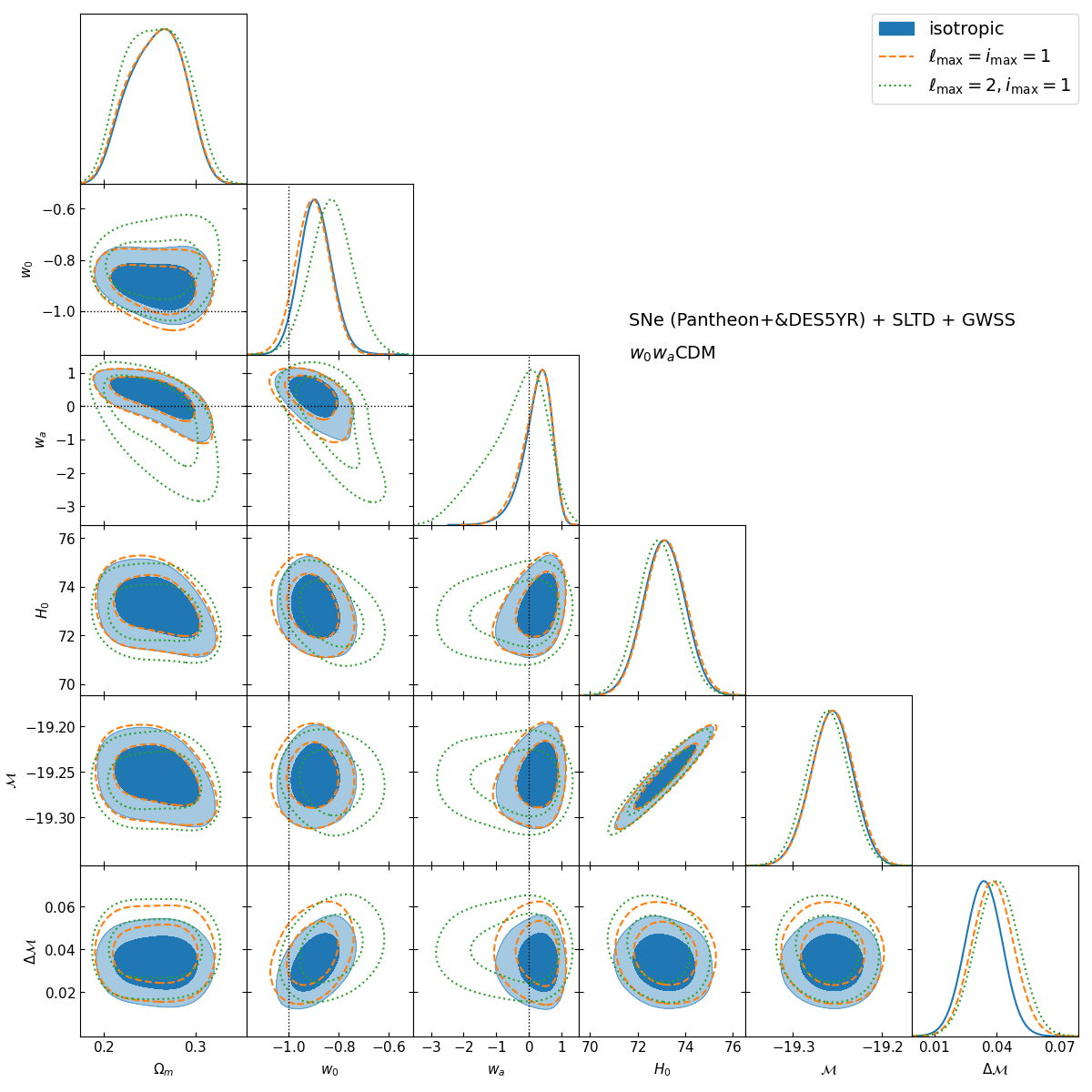}
    \caption{Marginalized 68.3\% and 95.4\% confidence-level constraints on cosmological parameters.\label{fig:cosmo}}
\end{figure}

As more data accumulates, cosmological information does not seem to converge towards $\Lambda$CDM. Both the isotropic $w_0w_a$CDM and the isotropic PAge remain competitive in terms of low AICs, though BIC criterion tends to favor the isotropic $\Lambda$CDM. Moreover, the inferred cosmological parameters, some of which are shown in Figure~\ref{fig:cosmo}, are not fully consistent with their values constrained by CMB+$\Lambda$CDM.  For isotropic $\Lambda$CDM model we obtain $H_0 = 72.78\pm 0.78\,\mathrm{km/s/Mpc}$, in $5.8\sigma$ tension with the Planck value $67.4\pm 0.5\,\mathrm{km/s/Mpc}$~\cite{Planck2018params}. Compared to the SH0ES result $H_0=73.04\pm 1.04\,\mathrm{km/s/Mpc}$, our tighter constraint on $H_0$ is mainly due to inclusion of the SLTD data which also measures $H_0$. For isotropic $w_0w_a$ model we find $w_0=-0.89\pm 0.07$ and $w_a = 0.26\pm 0.45$, i.e., no evidence for the recently claimed dynamical dark energy~\cite{DESI_2024, DESI_2025, DES_BAO_2025}. However, due to the controversial joint usage of different SN catalogs, these cosmological results based on the constant $\Delta\mathcal{M}$ model should be quoted with caution. In the anisotropic case, we find significant degeneracy between $w_a$ and the quadrupole coefficients, which substantially increases the uncertainty of $w_a$. The strongest degeneracy is between $w_a$ and $C_{i=1, \ell=2, m=0}$, with a statistical correlation $\mathrm{Corr}(w_a, C_{120}) = 0.69$. This degeneracy, along with others among the orthogonal mode coefficients, arises because the limited and non-uniform spatial distribution of the supernova sample leads to incomplete measurements of the orthogonal modes—particularly those at high resolution.

\begin{figure}
  \centering
  \includegraphics[width=0.5\linewidth]{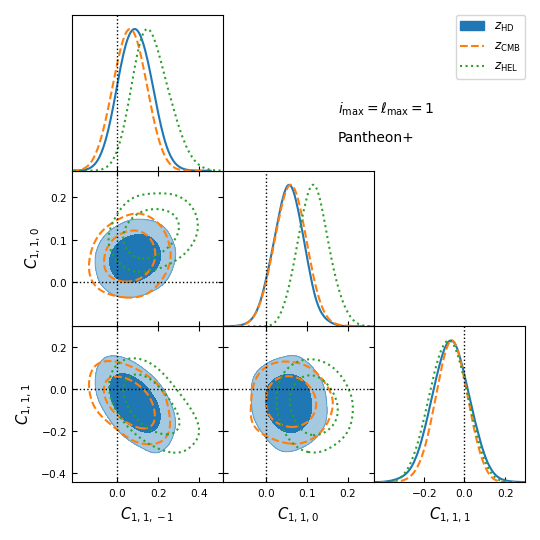}%
  \includegraphics[width=0.5\linewidth]{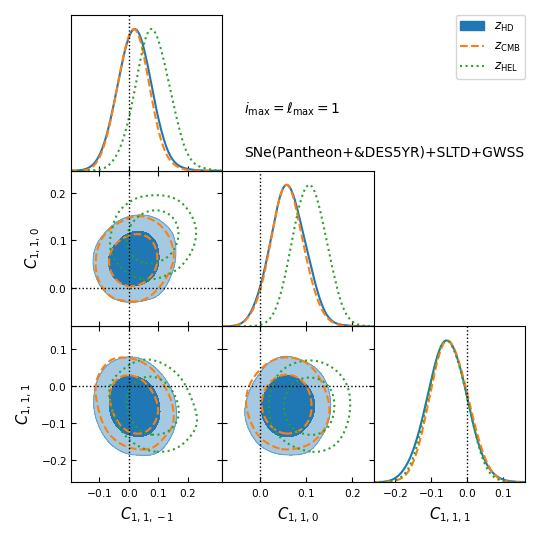}  
  \caption{Analyses with solar-frame redshift $z_{\rm HEL}$, CMB-frame redshift $z_{\rm CMB}$ and cosmological redshift $z_{\rm HD}$. Left panel: Pantheon+; right panel: SNe(Pantheon+\&DES5YR)+SLTD+GWSS. \label{fig:pec}}
\end{figure}

We have qualitatively estimated that our analysis is not sensitive to the peculiar velocity correction—that is, the difference between $z_{\rm HD}$ and $z_{\rm CMB}$. We now examine this issue more thoroughly through quantitative means. To do so, we repeat the analysis by replacing the cosmological redshift $z_{\rm HD}$ in the likelihood with both $z_{\rm CMB}$ and $z_{\rm HEL}$. As shown in Figure~\ref{fig:pec}, while using $z_{\rm HEL}$ instead of $z_{\rm HD}$ clearly reveals a dipole-like anisotropy caused by the solar motion relative to the CMB frame, the results obtained with $z_{\rm CMB}$ are nearly identical to those using $z_{\rm HD}$. Nevertheless, this does not imply that the bulk-flow velocity in the local universe is substantially smaller than the solar peculiar motion. The solar motion has a coherent effect that influences the redshifts of all supernovae up to several $\mathrm{Gpc}$ away, whereas bulk flows predominantly affect objects within approximately $250\,\mathrm{Mpc}$. As figure~\ref{fig:basis} shows, our model is not designed to detect inhomogeneity/anisotropy on sub-$\mathrm{Gpc}$ scales. Therefore, the insensitivity of our results to peculiar velocity correction is a prior choice rather than physical constraint.

\section{Discussion and conclusions \label{sec:con}}

In this work, we apply a full 3D expansion of anisotropic modes to test the cosmological principle.  Applying the new method to the Pantheon+ supernova catalog, and to Pantheon+ combined with DES5YR SNe, SLTD, and GWSS, we find no hint of anisotropy on large scales. These results deny the global hemisphere asymmetry found in Ref.~\cite{Hu:2024qnx}. As we have emphasized in the introduction, a seemingly global asymmetry found with the hemisphere comparison method could be actually local, as the chosen parameter ($H_0$) may only be sensitive to the local (distance-ladder) data. Therefore, the hemisphere asymmetry identified in Ref.~\cite{Hu:2024qnx} may be another manifestation of the local dipole-like anisotropy at $z\lesssim 0.1$~\cite{Watkins:2023rll, Tang_2023, Sorrenti_2023, Sah_2024, Sorrenti_2025}.

For the anisotropic model, both the AIC and BIC values increase with the number of radial modes ($i_{\max}$), indicating that introducing more complex radial dependencies ($i_{\max} \ge 2$) constitutes an over-parameterization given the currently available data. This reflects the observation that the investigated data do not prefer any large-scale anisotropic deviations, even when radially varying models as those shown here are considered.


Due to the limited size of current datasets, our analysis expands the Hubble residuals only to low resolutions ($i_{\max}, \ell_{\max} \sim$ a few). This construction makes it a global test of the cosmological principle on the largest accessible scales ($\sim\mathrm{Gpc}$), inherently smoothing over smaller-scale fluctuations. Consequently, our work does not capture the local ($z \lesssim 0.1$) anisotropy reported by other studies~\cite{Tang_2023, Sorrenti_2023, Cowell_2023, Dhawan_2023, Sah_2024, Sorrenti_2025}. Their findings, derived from methods targeting local bulk flows, probe a different physical regime—the dynamics of our immediate cosmic neighborhood. Thus, our null evidence for large-scale anisotropy is complementary to, rather than in conflict with, their results. Future data releases will enable higher-resolution expansions, bridging these scale-dependent tests.

\acknowledgments

This work is supported by the National key R\&D Program of China (Grant No. 2020YFC2201600), the National Natural Science Foundation of China (NSFC) under its Key Program (Grant No. 12533002) and General Program (Grant No. 12073088), and National SKA Program of China No. 2020SKA0110402.

\newpage{}

\begin{thebibliography}{10}

\bibitem{Milne_1932}
E.A.~{Milne}, \emph{{World Structure and the Expansion of the Universe}},
  \href{https://doi.org/10.1038/130009a0}{\emph{\nat} {\bfseries 130} (1932)
  9}.

\bibitem{Aluri:2022hzs}
P.K.~Aluri et~al., \emph{{Is the observable Universe consistent with the
  cosmological principle?}},
  \href{https://doi.org/10.1088/1361-6382/acbefc}{\emph{Class. Quant. Grav.}
  {\bfseries 40} (2023) 094001}
  [\href{https://arxiv.org/abs/2207.05765}{{\ttfamily 2207.05765}}].

\bibitem{Penzias_1965}
A.A.~{Penzias} and R.W.~{Wilson}, \emph{{A Measurement of Excess Antenna
  Temperature at 4080 Mc/s.}},
  \href{https://doi.org/10.1086/148307}{\emph{\apj} {\bfseries 142} (1965)
  419}.

\bibitem{Wilson_1967}
R.W.~{Wilson} and A.A.~{Penzias}, \emph{{Isotropy of Cosmic Background
  Radiation at 4080 Megahertz}},
  \href{https://doi.org/10.1126/science.156.3778.1100}{\emph{Science}
  {\bfseries 156} (1967) 1100}.

\bibitem{Partridge_1967}
R.B.~{Partridge} and D.T.~{Wilkinson}, \emph{{Isotropy and Homogeneity of the
  Universe from Measurements of the Cosmic Microwave Background}},
  \href{https://doi.org/10.1103/PhysRevLett.18.557}{\emph{\prl} {\bfseries 18}
  (1967) 557}.

\bibitem{Ryle_1968}
M.~{Ryle}, \emph{{The Counts of Radio Sources}},
  \href{https://doi.org/10.1146/annurev.aa.06.090168.001341}{\emph{\araa}
  {\bfseries 6} (1968) 249}.

\bibitem{Ellis_1984}
G.F.R.~{Ellis} and J.E.~{Baldwin}, \emph{{On the expected anisotropy of radio
  source counts}}, \href{https://doi.org/10.1093/mnras/206.2.377}{\emph{\mnras}
  {\bfseries 206} (1984) 377}.

\bibitem{Singal_2011}
A.K.~{Singal}, \emph{{Large Peculiar Motion of the Solar System from the Dipole
  Anisotropy in Sky Brightness due to Distant Radio Sources}},
  \href{https://doi.org/10.1088/2041-8205/742/2/L23}{\emph{\apjl} {\bfseries
  742} (2011) L23} [\href{https://arxiv.org/abs/1110.6260}{{\ttfamily
  1110.6260}}].

\bibitem{Rubart_2013}
M.~{Rubart} and D.J.~{Schwarz}, \emph{{Cosmic radio dipole from NVSS and
  WENSS}}, \href{https://doi.org/10.1051/0004-6361/201321215}{\emph{\aap}
  {\bfseries 555} (2013) A117}
  [\href{https://arxiv.org/abs/1301.5559}{{\ttfamily 1301.5559}}].

\bibitem{Tiwari_2016}
P.~{Tiwari} and A.~{Nusser}, \emph{{Revisiting the NVSS number count dipole}},
  \href{https://doi.org/10.1088/1475-7516/2016/03/062}{\emph{\jcap} {\bfseries
  2016} (2016) 062} [\href{https://arxiv.org/abs/1509.02532}{{\ttfamily
  1509.02532}}].

\bibitem{Bengaly_2018}
C.A.P.~{Bengaly}, R.~{Maartens} and M.G.~{Santos}, \emph{{Probing the
  Cosmological Principle in the counts of radio galaxies at different
  frequencies}},
  \href{https://doi.org/10.1088/1475-7516/2018/04/031}{\emph{\jcap} {\bfseries
  2018} (2018) 031} [\href{https://arxiv.org/abs/1710.08804}{{\ttfamily
  1710.08804}}].

\bibitem{Singal_2019}
A.K.~{Singal}, \emph{{Large disparity in cosmic reference frames determined
  from the sky distributions of radio sources and the microwave background
  radiation}}, \href{https://doi.org/10.1103/PhysRevD.100.063501}{\emph{\prd}
  {\bfseries 100} (2019) 063501}
  [\href{https://arxiv.org/abs/1904.11362}{{\ttfamily 1904.11362}}].

\bibitem{Siewert_2021}
T.M.~{Siewert}, M.~{Schmidt-Rubart} and D.J.~{Schwarz}, \emph{{Cosmic radio
  dipole: Estimators and frequency dependence}},
  \href{https://doi.org/10.1051/0004-6361/202039840}{\emph{\aap} {\bfseries
  653} (2021) A9} [\href{https://arxiv.org/abs/2010.08366}{{\ttfamily
  2010.08366}}].

\bibitem{Singal_2023}
A.K.~{Singal}, \emph{{Discordance of dipole asymmetries seen in recent large
  radio surveys with the cosmological principle}},
  \href{https://doi.org/10.1093/mnras/stad2161}{\emph{\mnras} {\bfseries 524}
  (2023) 3636} [\href{https://arxiv.org/abs/2303.05141}{{\ttfamily
  2303.05141}}].

\bibitem{Secrest_2021}
N.J.~{Secrest}, S.~{von Hausegger}, M.~{Rameez}, R.~{Mohayaee}, S.~{Sarkar} and
  J.~{Colin}, \emph{{A Test of the Cosmological Principle with Quasars}},
  \href{https://doi.org/10.3847/2041-8213/abdd40}{\emph{\apjl} {\bfseries 908}
  (2021) L51} [\href{https://arxiv.org/abs/2009.14826}{{\ttfamily
  2009.14826}}].

\bibitem{Secrest_2022}
N.J.~{Secrest}, S.~{von Hausegger}, M.~{Rameez}, R.~{Mohayaee} and S.~{Sarkar},
  \emph{{A Challenge to the Standard Cosmological Model}},
  \href{https://doi.org/10.3847/2041-8213/ac88c0}{\emph{\apjl} {\bfseries 937}
  (2022) L31} [\href{https://arxiv.org/abs/2206.05624}{{\ttfamily
  2206.05624}}].

\bibitem{Dam_2023}
L.~{Dam}, G.F.~{Lewis} and B.J.~{Brewer}, \emph{{Testing the cosmological
  principle with CatWISE quasars: a bayesian analysis of the number-count
  dipole}}, \href{https://doi.org/10.1093/mnras/stad2322}{\emph{\mnras}
  {\bfseries 525} (2023) 231}
  [\href{https://arxiv.org/abs/2212.07733}{{\ttfamily 2212.07733}}].

\bibitem{Wagenveld_2024}
J.D.~{Wagenveld}, H.R.~{Kl{\"o}ckner}, N.~{Gupta}, S.~{Sekhar},
  P.~{Jagannathan}, P.P.~{Deka} et~al., \emph{{The MeerKAT Absorption Line
  Survey Data Release 2: Wideband continuum catalogues and a measurement of the
  cosmic radio dipole}},
  \href{https://doi.org/10.1051/0004-6361/202450291}{\emph{\aap} {\bfseries
  690} (2024) A163} [\href{https://arxiv.org/abs/2408.16619}{{\ttfamily
  2408.16619}}].

\bibitem{Watkins:2008hf}
R.~Watkins, H.A.~Feldman and M.J.~Hudson, \emph{{Consistently Large Cosmic
  Flows on Scales of 100 Mpc/h: a Challenge for the Standard LCDM Cosmology}},
  \href{https://doi.org/10.1111/j.1365-2966.2008.14089.x}{\emph{\mnras}
  {\bfseries 392} (2009) 743}
  [\href{https://arxiv.org/abs/0809.4041}{{\ttfamily 0809.4041}}].

\bibitem{Feldman:2009es}
H.A.~Feldman, R.~Watkins and M.J.~Hudson, \emph{{Cosmic Flows on 100 Mpc/h
  Scales: Standardized Minimum Variance Bulk Flow, Shear and Octupole
  Moments}},
  \href{https://doi.org/10.1111/j.1365-2966.2010.17052.x}{\emph{\mnras}
  {\bfseries 407} (2010) 2328}
  [\href{https://arxiv.org/abs/0911.5516}{{\ttfamily 0911.5516}}].

\bibitem{Watkins:2023rll}
R.~Watkins, T.~Allen, C.J.~Bradford, A.~Ramon, A.~Walker, H.A.~Feldman et~al.,
  \emph{{Analysing the large-scale bulk flow using cosmicflows4: increasing
  tension with the standard cosmological model}},
  \href{https://doi.org/10.1093/mnras/stad1984}{\emph{\mnras} {\bfseries 524}
  (2023) 1885} [\href{https://arxiv.org/abs/2302.02028}{{\ttfamily
  2302.02028}}].

\bibitem{Sorrenti_2023}
F.~{Sorrenti}, R.~{Durrer} and M.~{Kunz}, \emph{{The dipole of the
  Pantheon+SH0ES data}},
  \href{https://doi.org/10.1088/1475-7516/2023/11/054}{\emph{\jcap} {\bfseries
  2023} (2023) 054} [\href{https://arxiv.org/abs/2212.10328}{{\ttfamily
  2212.10328}}].

\bibitem{Tang_2023}
L.~{Tang}, H.-N.~{Lin}, L.~{Liu} and X.~{Li}, \emph{{Consistency of Pantheon+
  supernovae with a large-scale isotropic universe}},
  \href{https://doi.org/10.1088/1674-1137/acfaf0}{\emph{Chinese Physics C}
  {\bfseries 47} (2023) 125101}
  [\href{https://arxiv.org/abs/2309.11320}{{\ttfamily 2309.11320}}].

\bibitem{Sah_2024}
A.~{Sah}, M.~{Rameez}, S.~{Sarkar} and C.G.~{Tsagas}, \emph{{Anisotropy in
  Pantheon+ supernovae}},
  \href{https://doi.org/10.1140/epjc/s10052-025-14222-w}{\emph{European
  Physical Journal C} {\bfseries 85} (2025) 596}
  [\href{https://arxiv.org/abs/2411.10838}{{\ttfamily 2411.10838}}].

\bibitem{Cowell_2023}
J.A.~{Cowell}, S.~{Dhawan} and H.J.~{Macpherson}, \emph{{Potential signature of
  a quadrupolar hubble expansion in Pantheon+supernovae}},
  \href{https://doi.org/10.1093/mnras/stad2788}{\emph{\mnras} {\bfseries 526}
  (2023) 1482} [\href{https://arxiv.org/abs/2212.13569}{{\ttfamily
  2212.13569}}].

\bibitem{Dhawan_2023}
S.~{Dhawan}, A.~{Borderies}, H.J.~{Macpherson} and A.~{Heinesen}, \emph{{The
  quadrupole in the local Hubble parameter: first constraints using Type Ia
  supernova data and forecasts for future surveys}},
  \href{https://doi.org/10.1093/mnras/stac3812}{\emph{\mnras} {\bfseries 519}
  (2023) 4841} [\href{https://arxiv.org/abs/2205.12692}{{\ttfamily
  2205.12692}}].

\bibitem{Sorrenti_2025}
F.~{Sorrenti}, R.~{Durrer} and M.~{Kunz}, \emph{{The low multipoles in the
  Pantheon+SH0ES data}},
  \href{https://doi.org/10.1088/1475-7516/2025/04/013}{\emph{\jcap} {\bfseries
  2025} (2025) 013} [\href{https://arxiv.org/abs/2403.17741}{{\ttfamily
  2403.17741}}].

\bibitem{Shurtleff:2013sua}
R.~Shurtleff, \emph{{A Large Scale Pattern from Optical Quasar Polarization
  Vectors}},  \href{https://arxiv.org/abs/1311.6118}{{\ttfamily 1311.6118}}.

\bibitem{Pelgrims_2015}
V.~{Pelgrims} and D.~{Hutsem{\'e}kers}, \emph{{Polarization alignments of
  quasars from the JVAS/CLASS 8.4-GHz surveys}},
  \href{https://doi.org/10.1093/mnras/stv917}{\emph{\mnras} {\bfseries 450}
  (2015) 4161} [\href{https://arxiv.org/abs/1503.03482}{{\ttfamily
  1503.03482}}].

\bibitem{Shamir_2025}
L.~{Shamir}, \emph{{The distribution of galaxy rotation in JWST Advanced Deep
  Extragalactic Survey}},
  \href{https://doi.org/10.1093/mnras/staf292}{\emph{\mnras} {\bfseries 538}
  (2025) 76} [\href{https://arxiv.org/abs/2502.18781}{{\ttfamily 2502.18781}}].

\bibitem{Huang_2022b}
Z.~{Huang}, \emph{{The Galactic Interstellar Medium Has a Preferred Handedness
  of Magnetic Misalignment}},
  \href{https://doi.org/10.3390/universe8080423}{\emph{Universe} {\bfseries 8}
  (2022) 423} [\href{https://arxiv.org/abs/2206.15375}{{\ttfamily
  2206.15375}}].

\bibitem{Deng:2018jrp}
H.-K.~Deng and H.~Wei, \emph{{Null signal for the cosmic anisotropy in the
  Pantheon supernovae data}},
  \href{https://doi.org/10.1140/epjc/s10052-018-6159-4}{\emph{Eur. Phys. J. C}
  {\bfseries 78} (2018) 755}
  [\href{https://arxiv.org/abs/1806.02773}{{\ttfamily 1806.02773}}].

\bibitem{Zhao:2019azy}
D.~Zhao, Y.~Zhou and Z.~Chang, \emph{{Anisotropy of the Universe via the
  Pantheon supernovae sample revisited}},
  \href{https://doi.org/10.1093/mnras/stz1259}{\emph{\mnras} {\bfseries 486}
  (2019) 5679} [\href{https://arxiv.org/abs/1903.12401}{{\ttfamily
  1903.12401}}].

\bibitem{Hu_2024}
J.P.~{Hu}, Y.Y.~{Wang}, J.~{Hu} and F.Y.~{Wang}, \emph{{Testing the
  cosmological principle with the Pantheon+ sample and the region-fitting
  method}}, \href{https://doi.org/10.1051/0004-6361/202347121}{\emph{\aap}
  {\bfseries 681} (2024) A88}
  [\href{https://arxiv.org/abs/2310.11727}{{\ttfamily 2310.11727}}].

\bibitem{Hu:2024qnx}
J.~Hu, J.~Hu, X.~Jia, B.~Gao and F.~Wang, \emph{{Testing cosmic anisotropy with
  Pad\'e approximations and the latest Pantheon+ sample}},
  \href{https://doi.org/10.1051/0004-6361/202450342}{\emph{Astron. Astrophys.}
  {\bfseries 689} (2024) A215}
  [\href{https://arxiv.org/abs/2406.14827}{{\ttfamily 2406.14827}}].

\bibitem{Yang_2025}
Z.-F.~{Yang}, D.-W.~{Yao}, M.~{Le Delliou} and K.~{Wang},
  \emph{{Model-independent test of the cosmic anisotropy with inverse distance
  ladder}},
  \href{https://doi.org/10.1140/epjc/s10052-025-13994-5}{\emph{European
  Physical Journal C} {\bfseries 85} (2025) 339}
  [\href{https://arxiv.org/abs/2407.19278}{{\ttfamily 2407.19278}}].

\bibitem{Schwarz_2007}
D.J.~{Schwarz} and B.~{Weinhorst}, \emph{{(An)isotropy of the Hubble diagram:
  comparing hemispheres}},
  \href{https://doi.org/10.1051/0004-6361:20077998}{\emph{\aap} {\bfseries 474}
  (2007) 717} [\href{https://arxiv.org/abs/0706.0165}{{\ttfamily 0706.0165}}].

\bibitem{Antoniou_2010}
I.~Antoniou and L.~Perivolaropoulos, \emph{Searching for a cosmological
  preferred axis: Union2 data analysis and comparison with other probes},
  \href{https://doi.org/10.1088/1475-7516/2010/12/012}{\emph{Journal of
  Cosmology and Astroparticle Physics} {\bfseries 2010} (2010) 012–012}.

\bibitem{Cai:2011xs}
R.-G.~Cai and Z.-L.~Tuo, \emph{{Direction Dependence of the Deceleration
  Parameter}}, \href{https://doi.org/10.1088/1475-7516/2012/02/004}{\emph{JCAP}
  {\bfseries 02} (2012) 004} [\href{https://arxiv.org/abs/1109.0941}{{\ttfamily
  1109.0941}}].

\bibitem{Yang:2013gea}
X.~Yang, F.Y.~Wang and Z.~Chu, \emph{{Searching for a preferred direction with
  Union2.1 data}}, \href{https://doi.org/10.1093/mnras/stt2015}{\emph{\mnras}
  {\bfseries 437} (2014) 1840}
  [\href{https://arxiv.org/abs/1310.5211}{{\ttfamily 1310.5211}}].

\bibitem{Hu:2020mzd}
J.P.~Hu, Y.Y.~Wang and F.Y.~Wang, \emph{{Testing cosmic anisotropy with
  Pantheon sample and quasars at high redshifts}},
  \href{https://doi.org/10.1051/0004-6361/202038541}{\emph{Astron. Astrophys.}
  {\bfseries 643} (2020) A93}
  [\href{https://arxiv.org/abs/2008.12439}{{\ttfamily 2008.12439}}].

\bibitem{Conville_2023}
R.~{Mc Conville} and E.~{{\'O} Colg{\'a}in}, \emph{{Anisotropic distance ladder
  in Pantheon+supernovae}},
  \href{https://doi.org/10.1103/PhysRevD.108.123533}{\emph{\prd} {\bfseries
  108} (2023) 123533} [\href{https://arxiv.org/abs/2304.02718}{{\ttfamily
  2304.02718}}].

\bibitem{Fixsen_2009}
D.J.~{Fixsen}, \emph{{The Temperature of the Cosmic Microwave Background}},
  \href{https://doi.org/10.1088/0004-637X/707/2/916}{\emph{\apj} {\bfseries
  707} (2009) 916} [\href{https://arxiv.org/abs/0911.1955}{{\ttfamily
  0911.1955}}].

\bibitem{Seager_1999}
S.~{Seager}, D.D.~{Sasselov} and D.~{Scott}, \emph{{A New Calculation of the
  Recombination Epoch}}, \href{https://doi.org/10.1086/312250}{\emph{\apjl}
  {\bfseries 523} (1999) L1}
  [\href{https://arxiv.org/abs/astro-ph/9909275}{{\ttfamily
  astro-ph/9909275}}].

\bibitem{Chevallier:2000qy}
M.~Chevallier and D.~Polarski, \emph{{Accelerating universes with scaling dark
  matter}}, \href{https://doi.org/10.1142/S0218271801000822}{\emph{Int. J. Mod.
  Phys. D} {\bfseries 10} (2001) 213}
  [\href{https://arxiv.org/abs/gr-qc/0009008}{{\ttfamily gr-qc/0009008}}].

\bibitem{Linder:2002et}
E.V.~Linder, \emph{{Exploring the expansion history of the universe}},
  \href{https://doi.org/10.1103/PhysRevLett.90.091301}{\emph{Phys. Rev. Lett.}
  {\bfseries 90} (2003) 091301}
  [\href{https://arxiv.org/abs/astro-ph/0208512}{{\ttfamily
  astro-ph/0208512}}].

\bibitem{Huang_2020}
Z.~{Huang}, \emph{{Supernova Magnitude Evolution and PAge Approximation}},
  \href{https://doi.org/10.3847/2041-8213/ab8011}{\emph{\apjl} {\bfseries 892}
  (2020) L28} [\href{https://arxiv.org/abs/2001.06926}{{\ttfamily
  2001.06926}}].

\bibitem{Huang_2021}
L.~{Huang}, Z.-Q.~{Huang}, Z.-Y.~{Li} and H.~{Zhou}, \emph{{A more accurate
  Parameterization based on cosmic Age (MAPAge)}},
  \href{https://doi.org/10.1088/1674-4527/21/11/277}{\emph{Research in
  Astronomy and Astrophysics} {\bfseries 21} (2021) 277}
  [\href{https://arxiv.org/abs/2108.03959}{{\ttfamily 2108.03959}}].

\bibitem{Huang_2022a}
Z.~{Huang}, \emph{{Thawing k-essence dark energy in the PAge space}},
  \href{https://doi.org/10.1088/1572-9494/ac80ed}{\emph{Communications in
  Theoretical Physics} {\bfseries 74} (2022) 095404}
  [\href{https://arxiv.org/abs/2204.09713}{{\ttfamily 2204.09713}}].

\bibitem{Wang_2024}
J.~{Wang}, Z.~{Huang}, Y.~{Yao}, J.~{Liu}, L.~{Huang} and Y.~{Su}, \emph{{A
  PAge-like Unified Dark Fluid model}},
  \href{https://doi.org/10.1088/1475-7516/2024/09/053}{\emph{\jcap} {\bfseries
  2024} (2024) 053} [\href{https://arxiv.org/abs/2405.05798}{{\ttfamily
  2405.05798}}].

\bibitem{Su_2025}
Y.~{Su}, Z.~{Huang}, J.~{Wang}, Y.~{Yao} and J.~{Liu}, \emph{{The dark side of
  the universe may be more harmonic than we thought}},
  \href{https://doi.org/10.48550/arXiv.2504.00536}{\emph{arXiv e-prints} (2025)
  arXiv:2504.00536} [\href{https://arxiv.org/abs/2504.00536}{{\ttfamily
  2504.00536}}].

\bibitem{Brout_2022}
D.~{Brout}, D.~{Scolnic}, B.~{Popovic}, A.G.~{Riess} et~al., \emph{{The
  Pantheon+ Analysis: Cosmological Constraints}},
  \href{https://doi.org/10.3847/1538-4357/ac8e04}{\emph{\apj} {\bfseries 938}
  (2022) 110} [\href{https://arxiv.org/abs/2202.04077}{{\ttfamily
  2202.04077}}].

\bibitem{Mazurenko_2024}
S.~{Mazurenko}, I.~{Banik}, P.~{Kroupa} and M.~{Haslbauer}, \emph{{A
  simultaneous solution to the Hubble tension and observed bulk flow within 250
  h$^{-1}$ Mpc}}, \href{https://doi.org/10.1093/mnras/stad3357}{\emph{\mnras}
  {\bfseries 527} (2024) 4388}
  [\href{https://arxiv.org/abs/2311.17988}{{\ttfamily 2311.17988}}].

\bibitem{Koksbang_2024}
S.M.~{Koksbang}, \emph{{Testing inhomogeneous cosmography in our cosmic
  neighborhood using CosmicFlows-4}},
  \href{https://doi.org/10.1103/6z7w-47rc}{\emph{\prd} {\bfseries 111} (2025)
  123516} [\href{https://arxiv.org/abs/2412.12637}{{\ttfamily 2412.12637}}].

\bibitem{Tzartinoglou_2024}
A.~{Tzartinoglou} and C.G.~{Tsagas}, \emph{{The deceleration parameter in
  perturbed Bianchi universes with a peculiar-velocity ``tilt''}},
  \href{https://doi.org/10.1140/epjc/s10052-024-13358-5}{\emph{European
  Physical Journal C} {\bfseries 84} (2024) 1061}
  [\href{https://arxiv.org/abs/2405.17592}{{\ttfamily 2405.17592}}].

\bibitem{Li_2020}
M.~{Li}, H.~{Rao} and D.~{Zhao}, \emph{{A simple parity violating gravity model
  without ghost instability}},
  \href{https://doi.org/10.1088/1475-7516/2020/11/023}{\emph{\jcap} {\bfseries
  2020} (2020) 023} [\href{https://arxiv.org/abs/2007.08038}{{\ttfamily
  2007.08038}}].

\bibitem{Li_2021}
M.~{Li}, H.~{Rao} and Y.~{Tong}, \emph{{Revisiting a parity violating gravity
  model without ghost instability: Local Lorentz covariance}},
  \href{https://doi.org/10.1103/PhysRevD.104.084077}{\emph{\prd} {\bfseries
  104} (2021) 084077} [\href{https://arxiv.org/abs/2104.05917}{{\ttfamily
  2104.05917}}].

\bibitem{Cai_2022}
R.-G.~{Cai}, C.~{Fu} and W.-W.~{Yu}, \emph{{Parity violation in stochastic
  gravitational wave background from inflation in Nieh-Yan modified
  teleparallel gravity}},
  \href{https://doi.org/10.1103/PhysRevD.105.103520}{\emph{\prd} {\bfseries
  105} (2022) 103520}.

\bibitem{Li_2022}
M.~{Li} and D.~{Zhao}, \emph{{A simple parity violating model in the symmetric
  teleparallel gravity and its cosmological perturbations}},
  \href{https://doi.org/10.1016/j.physletb.2022.136968}{\emph{Physics Letters
  B} {\bfseries 827} (2022) 136968}
  [\href{https://arxiv.org/abs/2108.01337}{{\ttfamily 2108.01337}}].

\bibitem{Li_2023}
M.~{Li} and H.~{Rao}, \emph{{Irregular universe in the Nieh-Yan modified
  teleparallel gravity}},
  \href{https://doi.org/10.1016/j.physletb.2023.137929}{\emph{Physics Letters
  B} {\bfseries 841} (2023) 137929}
  [\href{https://arxiv.org/abs/2301.02847}{{\ttfamily 2301.02847}}].

\bibitem{Zhu_2023}
M.~{Zhu} and Y.~{Cai}, \emph{{Parity-violation in bouncing cosmology}},
  \href{https://doi.org/10.1007/JHEP04(2023)095}{\emph{Journal of High Energy
  Physics} {\bfseries 2023} (2023) 95}
  [\href{https://arxiv.org/abs/2301.13502}{{\ttfamily 2301.13502}}].

\bibitem{Rao_2023}
H.~{Rao} and D.~{Zhao}, \emph{{Parity violating scalar-tensor model in
  teleparallel gravity and its cosmological application}},
  \href{https://doi.org/10.1007/JHEP08(2023)070}{\emph{Journal of High Energy
  Physics} {\bfseries 2023} (2023) 70}
  [\href{https://arxiv.org/abs/2304.07138}{{\ttfamily 2304.07138}}].

\bibitem{Jiang_2024}
Z.-W.~{Jiang}, Y.~{Cai}, F.~{Wang} and Y.-S.~{Piao}, \emph{{Parity-violating
  primordial gravitational waves from null energy condition violation}},
  \href{https://doi.org/10.1007/JHEP09(2024)067}{\emph{Journal of High Energy
  Physics} {\bfseries 2024} (2024) 67}
  [\href{https://arxiv.org/abs/2406.16549}{{\ttfamily 2406.16549}}].

\bibitem{Bond_2009}
J.R.~{Bond}, A.V.~{Frolov}, Z.~{Huang} and L.~{Kofman}, \emph{{Non-Gaussian
  Curvature Spikes from Chaotic Billiards in Inflation Preheating}},
  \href{https://doi.org/10.1103/PhysRevLett.103.071301}{\emph{\prl} {\bfseries
  103} (2009) 071301} [\href{https://arxiv.org/abs/0903.3407}{{\ttfamily
  0903.3407}}].

\bibitem{Huang_2019}
Z.~{Huang}, \emph{{High-redshift minihaloes from modulated preheating}},
  \href{https://doi.org/10.1103/PhysRevD.99.103537}{\emph{\prd} {\bfseries 99}
  (2019) 103537} [\href{https://arxiv.org/abs/1902.10096}{{\ttfamily
  1902.10096}}].

\bibitem{McInnes_2025}
B.~{McInnes}, \emph{{Reconciling Inflation with Hubble Anisotropies}},
  \href{https://doi.org/10.48550/arXiv.2508.06796}{\emph{arXiv e-prints} (2025)
  arXiv:2508.06796} [\href{https://arxiv.org/abs/2508.06796}{{\ttfamily
  2508.06796}}].

\bibitem{Huang_2025}
Z.~{Huang}, X.~{Ouyang}, Y.~{Cui}, J.~{Liu}, Y.~{Yao}, Z.~{Qiu} et~al.,
  \emph{{Curvature perturbations from kinetic preheating after
  {\ensuremath{\alpha}}-attractor inflation}},
  \href{https://doi.org/10.1103/djqm-5py3}{\emph{\prd} {\bfseries 112} (2025)
  043530} [\href{https://arxiv.org/abs/2408.14881}{{\ttfamily 2408.14881}}].

\bibitem{Wong_2020_H0LiCOW}
K.C.~{Wong}, S.H.~{Suyu}, G.C.F.~{Chen}, C.E.~{Rusu}, M.~{Millon}, D.~{Sluse}
  et~al., \emph{{H0LiCOW - XIII. A 2.4 per cent measurement of H$_{0}$ from
  lensed quasars: 5.3{\ensuremath{\sigma}} tension between early- and
  late-Universe probes}},
  \href{https://doi.org/10.1093/mnras/stz3094}{\emph{\mnras} {\bfseries 498}
  (2020) 1420} [\href{https://arxiv.org/abs/1907.04869}{{\ttfamily
  1907.04869}}].

\bibitem{Shajib_2020_STRIDES}
A.J.~{Shajib}, S.~{Birrer}, T.~{Treu}, A.~{Agnello}, E.J.~{Buckley-Geer},
  J.H.H.~{Chan} et~al., \emph{{STRIDES: a 3.9 per cent measurement of the
  Hubble constant from the strong lens system DES J0408-5354}},
  \href{https://doi.org/10.1093/mnras/staa828}{\emph{\mnras} {\bfseries 494}
  (2020) 6072} [\href{https://arxiv.org/abs/1910.06306}{{\ttfamily
  1910.06306}}].

\bibitem{Abbott_2017}
B.P.~{Abbott}, R.~{Abbott}, T.D.~{Abbott}, F.~{Acernese} et~al., \emph{{A
  gravitational-wave standard siren measurement of the Hubble constant}},
  \href{https://doi.org/10.1038/nature24471}{\emph{\nat} {\bfseries 551} (2017)
  85} [\href{https://arxiv.org/abs/1710.05835}{{\ttfamily 1710.05835}}].

\bibitem{DES_SN_2024}
{DES Collaboration}, T.M.C.~{Abbott}, M.~{Acevedo}, M.~{Aguena}, A.~{Alarcon}
  et~al., \emph{{The Dark Energy Survey: Cosmology Results with
  {\ensuremath{\sim}}1500 New High-redshift Type Ia Supernovae Using the Full 5
  yr Data Set}}, \href{https://doi.org/10.3847/2041-8213/ad6f9f}{\emph{\apjl}
  {\bfseries 973} (2024) L14}
  [\href{https://arxiv.org/abs/2401.02929}{{\ttfamily 2401.02929}}].

\bibitem{Suyu:2009by}
S.H.~Suyu, P.J.~Marshall, M.W.~Auger, S.~Hilbert, R.D.~Blandford,
  L.V.E.~Koopmans et~al., \emph{{Dissecting the Gravitational Lens B1608+656.
  II. Precision Measurements of the Hubble Constant, Spatial Curvature, and the
  Dark Energy Equation of State}},
  \href{https://doi.org/10.1088/0004-637X/711/1/201}{\emph{Astrophys. J.}
  {\bfseries 711} (2010) 201}
  [\href{https://arxiv.org/abs/0910.2773}{{\ttfamily 0910.2773}}].

\bibitem{Jee:2019hah}
I.~Jee, S.~Suyu, E.~Komatsu, C.D.~Fassnacht, S.~Hilbert and L.V.E.~Koopmans,
  \emph{{A measurement of the Hubble constant from angular diameter distances
  to two gravitational lenses}},
  \href{https://arxiv.org/abs/1909.06712}{{\ttfamily 1909.06712}}.

\bibitem{Suyu:2013kha}
S.H.~Suyu et~al., \emph{{Cosmology from gravitational lens time delays and
  Planck data}},
  \href{https://doi.org/10.1088/2041-8205/788/2/L35}{\emph{Astrophys. J. Lett.}
  {\bfseries 788} (2014) L35}
  [\href{https://arxiv.org/abs/1306.4732}{{\ttfamily 1306.4732}}].

\bibitem{H0LiCOW:2019xdh}
{\scshape H0LiCOW} collaboration, \emph{{A SHARP view of H0LiCOW: $H_{0}$ from
  three time-delay gravitational lens systems with adaptive optics imaging}},
  \href{https://doi.org/10.1093/mnras/stz2547}{\emph{\mnras} {\bfseries 490}
  (2019) 1743} [\href{https://arxiv.org/abs/1907.02533}{{\ttfamily
  1907.02533}}].

\bibitem{H0LiCOW:2016qrm}
{\scshape H0LiCOW} collaboration, \emph{{H0LiCOW \textendash{} IV. Lens mass
  model of HE 0435\ensuremath{-}1223 and blind measurement of its time-delay
  distance for cosmology}},
  \href{https://doi.org/10.1093/mnras/stw3077}{\emph{\mnras} {\bfseries 465}
  (2017) 4895} [\href{https://arxiv.org/abs/1607.01403}{{\ttfamily
  1607.01403}}].

\bibitem{H0LiCOW:2018tyj}
{\scshape H0LiCOW} collaboration, \emph{{H0LiCOW - IX. Cosmographic analysis of
  the doubly imaged quasar SDSS 1206+4332 and a new measurement of the Hubble
  constant}}, \href{https://doi.org/10.1093/mnras/stz200}{\emph{\mnras}
  {\bfseries 484} (2019) 4726}
  [\href{https://arxiv.org/abs/1809.01274}{{\ttfamily 1809.01274}}].

\bibitem{H0LiCOW:2019mdu}
{\scshape H0LiCOW} collaboration, \emph{{H0LiCOW XII. Lens mass model of
  WFI2033 \ensuremath{-} 4723 and blind measurement of its time-delay distance
  and H0}}, \href{https://doi.org/10.1093/mnras/stz3451}{\emph{\mnras}
  {\bfseries 498} (2020) 1440}
  [\href{https://arxiv.org/abs/1905.09338}{{\ttfamily 1905.09338}}].

\bibitem{Agnello:2017mwu}
A.~Agnello et~al., \emph{{Models of the strongly lensed quasar DES
  J0408\ensuremath{-}5354}},
  \href{https://doi.org/10.1093/mnras/stx2242}{\emph{\mnras} {\bfseries 472}
  (2017) 4038} [\href{https://arxiv.org/abs/1702.00406}{{\ttfamily
  1702.00406}}].

\bibitem{Kenworthy_2021}
W.D.~{Kenworthy}, D.O.~{Jones}, M.~{Dai}, R.~{Kessler}, D.~{Scolnic},
  D.~{Brout} et~al., \emph{{SALT3: An Improved Type Ia Supernova Model for
  Measuring Cosmic Distances}},
  \href{https://doi.org/10.3847/1538-4357/ac30d8}{\emph{\apj} {\bfseries 923}
  (2021) 265} [\href{https://arxiv.org/abs/2104.07795}{{\ttfamily
  2104.07795}}].

\bibitem{Guy_2007}
J.~{Guy}, P.~{Astier}, S.~{Baumont}, D.~{Hardin}, R.~{Pain}, N.~{Regnault}
  et~al., \emph{{SALT2: using distant supernovae to improve the use of type Ia
  supernovae as distance indicators}},
  \href{https://doi.org/10.1051/0004-6361:20066930}{\emph{\aap} {\bfseries 466}
  (2007) 11} [\href{https://arxiv.org/abs/astro-ph/0701828}{{\ttfamily
  astro-ph/0701828}}].

\bibitem{Efstathiou_2025}
G.~{Efstathiou}, \emph{{Evolving dark energy or supernovae systematics?}},
  \href{https://doi.org/10.1093/mnras/staf301}{\emph{\mnras} {\bfseries 538}
  (2025) 875} [\href{https://arxiv.org/abs/2408.07175}{{\ttfamily
  2408.07175}}].

\bibitem{Vincenzi_2025}
M.~{Vincenzi}, R.~{Kessler}, P.~{Shah} et~al., \emph{{Comparing the DES-SN5YR
  and Pantheon+ SN cosmology analyses: investigation based on 'evolving dark
  energy or supernovae systematics'?}},
  \href{https://doi.org/10.1093/mnras/staf943}{\emph{\mnras} {\bfseries 541}
  (2025) 2585} [\href{https://arxiv.org/abs/2501.06664}{{\ttfamily
  2501.06664}}].

\bibitem{Planck2018params}
{Planck Collaboration}, N.~{Aghanim}, Y.~{Akrami}, M.~{Ashdown} et~al.,
  \emph{{Planck 2018 results. VI. Cosmological parameters}},
  \href{https://doi.org/10.1051/0004-6361/201833910}{\emph{\aap} {\bfseries
  641} (2020) A6} [\href{https://arxiv.org/abs/1807.06209}{{\ttfamily
  1807.06209}}].

\bibitem{DESI_2024}
{\scshape DESI} collaboration, \emph{{DESI 2024 VI: cosmological constraints
  from the measurements of baryon acoustic oscillations}},
  \href{https://doi.org/10.1088/1475-7516/2025/02/021}{\emph{JCAP} {\bfseries
  02} (2025) 021} [\href{https://arxiv.org/abs/2404.03002}{{\ttfamily
  2404.03002}}].

\bibitem{DESI_2025}
M.~{Abdul Karim}, J.~{Aguilar}, S.~{Ahlen} et~al., \emph{{DESI DR2 results. II.
  Measurements of baryon acoustic oscillations and cosmological constraints}},
  \href{https://doi.org/10.1103/tr6y-kpc6}{\emph{\prd} {\bfseries 112} (2025)
  083515} [\href{https://arxiv.org/abs/2503.14738}{{\ttfamily 2503.14738}}].

\bibitem{DES_BAO_2025}
T.M.C.~{Abbott}, M.~{Acevedo}, M.~{Adamow} et~al., \emph{{Dark Energy Survey:
  implications for cosmological expansion models from the final DES Baryon
  Acoustic Oscillation and Supernova data}},
  \href{https://doi.org/10.48550/arXiv.2503.06712}{\emph{arXiv e-prints} (2025)
  arXiv:2503.06712} [\href{https://arxiv.org/abs/2503.06712}{{\ttfamily
  2503.06712}}].

\end{thebibliography}

\providecommand{\href}[2]{#2}\begingroup\raggedright\endgroup

\end{document}